# High-Sensitivity and Compact Time-domain Soil Moisture Sensor Using Dispersive Phase Shifter for Complex Permittivity Measurement

Rasool Keshavarz, and Negin Shariati, *IEEE Member*

*Abstract*— This paper presents a Time-Domain Transmissometry Soil Moisture Sensor (TDT-SMS) using a Dispersive Phase Shifter (DPS), consisting of an interdigital capacitor that is loaded with a stacked 4-turn Complementary Spiral Resonator (S4-CSR). Soil moisture measurement technique of the proposed sensor is based on the complex permittivity sensing property of a DPS in time domain. Soil relative permittivity which varies with its moisture content is measured by burying the DPS under a soil mass and changing its phase difference while excited with a 114 MHz sine wave (single tone). DPS output phase and magnitude are compared with the reference signal and measured with a phase/loss detector. The proposed sensor exhibits accuracy better than ±1.2% at the highest Volumetric Water Content (VWC=30%) for sandy-type soil. Precise design guide is developed and simulations are performed to achieve a highly sensitive sensor. The measurement results validate the accuracy of theoretical analysis and design procedure. Owning the advantages of low profile, low power consumption, and high sensitivity makes the proposed TDT-SMS a good candidate for precision farming and IoT systems.

*Index Terms*— Complex permittivity measurement, CSRR, metamaterial, phase shifter, soil moisture sensor, time-domain sensor, volumetric water content (VWC).

## I. INTRODUCTION

Soil moisture is a major contributing factor in environmental challenges such as climate change and natural disasters (e.g., flooding and landslides). Soil moisture plays a key role in precision agriculture due to its effect on drainage, infiltration, and fertilization, [1], [2]. Agriculture sensors are planted in the ground at different depths to measure and transmit soil properties to a central location. In this regard, Wireless Sensor Networks (WSN) and Internet of Things (IoT) offer an optimal solution by recording necessary information at a fast rate [3], [4].

Soil sensors can be categorized as electrical and electromagnetic, optical and radiometric, and electrochemical sensors [5]. Electrical and electromagnetic sensors are the most commonly used sensors in precision agriculture due to the information they can determine as well as their robustness and low-cost design compared to other soil sensors [6]. Some electromagnetic methods are based on relationship between the real part ($\varepsilon'$) of soil permittivity ($\varepsilon$) and volumetric water content (VWC) [7], [8]. However, $\varepsilon = \varepsilon' - j\varepsilon''$, and $\varepsilon''$ in moist soil is comparable with the $\varepsilon'$ value and it should be considered in the measurement setup [9]. For instance, $\varepsilon''$ in sandy soil changes from 0.05 to 3.5 for VWC variation of 0 to 30% at 130 MHz, and ignoring the imaginary part of the complex dielectric permittivity would degrade sensor calibration precision [10].

In terms of design methods, different electromagnetic (EM) techniques have been proposed to estimate the material under test (MUT) permittivity; time-domain reflectometry (TDR) [11], time-domain transmissometry (TDT), frequency domain reflectometry (FDR) [12], remote sensing, capacitance [13], and other methods [9], [14]. Some techniques require very complicated test sets or have a limited dynamic range. For example, FDR method estimates VWC based on a resonance frequency variation due to the soil dielectric properties. However, this method is based on a portable Vector Network Analyzer (VNA) technology and shares many of the cumbersome calibration and usage requirements for field measurements [15].

The principle of a typical TDT technique is based on the measurement of time interval that pulse travels from one starting point on the probe to the end, while the probe is surrounded by MUT (moist soil) [16]. However, when MUT is considered dispersive, its permittivity or permeability are functions of frequency ($\varepsilon(\omega), \mu(\omega)$). Hence, the group velocity in dispersive material is a function of frequency [17]. This leads to inaccurate measurement of the rising/falling edges of the traveling pulse [18] which is the main drawback of typical time-domain sensors. Since moist soil is like a dispersive media, different dielectric dispersion models like Debye 1st order, Debye 2nd order, Drude, Lorentz, etc. are considered to model dispersion in the moist soil [18]. Therefore, the traveling pulse along the embedded microstrip transmission line (EMTL) in moist soil is degraded, leading to imprecise measurement of time-domain sensors which are excited by pulse. Additionally, pulse distortion increases in inhomogeneous measurement areas which include multiple types of soils or non-uniform VWC.

The penetration of microwave signals into a MUT is defined by the skin depth coefficient, which is a function of its dielectric properties and frequency [19]. By decreasing the operational frequency band of the sensor, the skin depth increases and leads to a larger Volume Under Test (VUT) measurement. Hence, the sensor covers a greater testing area (large soil volume) which is suitable for decreasing the number of sensors in a wireless sensor network constellation. However, low-frequency sensors are bulky and have their limitations in use, implementation, maintenance, and transportation. Therefore, there is a tradeoff between choosing an optimal frequency and VUT value.

Metamaterial Transmission Lines (TLs) have been known to miniaturize microwave devices like filters [20], [21], couplers, and other components [22], [23]. Several types of resonators are used in the body of metamaterial TLs like split-ring resonator (SRR) and its complement, CSRR. There are several configurations of SRR and CSRR such as edge-coupled SRR (EC-SRR), broadside-coupled SRR (BC-SRR), nonbianisotropic SRR (NB-SRR), double-split SRR (2-SRR), and two-turn spiral resonator (2-SR) which can be used in microstrip transmission lines [24].

RF and Communication Technologies (RFCT) research laboratory, University of Technology Sydney, Ultimo, NSW 2007, Australia, e-mail: Rasool.Keshavarz@uts.edu.au; Negin.Shariati@uts.edu.au

In this paper, we design, simulate, test, and implement a compact Time-Domain Transmissometry Soil Moisture Sensor (TDT-SMS) based on sine wave (single tone) excitation. The proposed method, unlike pulse sensors, withstands the dispersive property of moist soil and improves the measurement accuracy. The proposed TDT-SMS consists of three main parts: reference oscillator, Dispersive Phase Shifter (DPS), and phase/loss detector. In the proposed method, DPS input port is excited by a sine wave (reference oscillator), while MUT (moist soil) is poured on the DPS top layer. Hence, the phase and magnitude differences between DPS output and reference signal indicate the real and imaginary parts of MUT permittivity. Moreover, a stacked four-turn complementary spiral resonator (S4-CSR) structure is used on the bottom layer of DPS to miniaturize its size and realize a large phase shift in a small length. Further, on the top layer of the proposed DPS, an interdigital capacitor is used to achieve a bandpass property in the operational frequency band (114 MHz). Finally, a commercial compact gain/loss detector (AD8302) is used to detect the phase difference and loss, which can be measured using a simple multimeter in an embedded scenario for industrial applications. In the proposed technique, the sensor is implemented without using any measurement equipment like oscilloscope, spectrum analyzer, etc. This method is capable to be used in embedded sensor scenarios and massive WSNs in field measurements.

Major contributions of this paper are summarized as follows:
- This work presents a TDT-SMS that is excited with a sine-wave instead of pulse. This technique improves the sensor accuracy compared to conventional methods.
- The proposed sensor measures the real and imaginary parts of a MUT (moist soil), simultaneously which enhance accuracy of the sensor calibration procedure.
- For the first time, a stacked 4-CSR in a DPS structure is designed, and the effect of equivalent circuit model components on the phase difference value is analyzed.
- In order to cover large VUT and also to achieve a compact structure, the proposed sensor is designed at low-frequency band (114 MHz), while the size is miniaturized using S4-CSR resonator. Theoretical analysis and investigation of commercialized products support the choice of the proposed operational frequency band.
- From practical design and economic perspectives, achieving a large phase difference in a compact structure and measuring the phase difference and loss using a compact detector, leads to a miniaturized and low-cost soil sensor. Hence, the proposed sensor is a good candidate for integrating into various industrial sensors and IoT systems.
- The design guide procedure for other soil types or arbitrary MUT and operational frequency bands are presented and equations are derived. Therefore, the proposed technique can be generalized and extended to other applications, as it has the capability to measure complex permittivity of different MUTs.
- The proposed sensor consists of two main parts: Dispersive Phase Shifter (DPS) as a passive structure and a phase/loss detector (AD8302) which consumes low DC power (<66 mW). Since the sensing process for soil moisture measurement takes short time (less than 1 sec) and repeats sometimes during a day, the proposed system acts as a low power sensor which is suitable for field measurement in massive wireless sensor networks in precision farming or other practical scenarios.

The organization of this paper is as follows: Sensor design and methodology are presented in Section II. The proposed TDT-SMS performance is validated by analytical, simulation, and measurement results in Section III. Finally, conclusions are provided in Section IV.

## II. SENSOR DESIGN AND METHODOLOGY

Typical time-domain soil sensors (TDR or TDT) provide an estimate of moisture content by measuring the relative permittivity in response to the soil VWC, which is determined by the time an electromagnetic pulse travels in soil. The main difference between TDR and TDT is that the latter measures the time of transmission and not reflection, as TDR [16]. In this section, firstly, we simulate the effect of moist soil as a dispersive media on the pulse distortion in typical time-domain sensors and compare it with the proposed sine wave excitation technique. Then, the theory and design guide of a DPS as the main part of the proposed TDT-SMS will be investigated.

### A. Time-Domain Sensors in Dispersive Media

In dispersive media, the permittivity or permeability of MUT are functions of frequency ($\varepsilon(\omega)$, $\mu(\omega)$). Dispersion leads to a pulse distortion which can be intensified when MUT is inhomogeneous and encompasses multiple soil types and non-uniform VWC. Figure 1 exhibits an inhomogeneous media which consists of five different materials. Moreover, since the permittivity of each material depends on frequency, this structure is a dispersive inhomogeneous media. Figure 2 exhibits CST simulation results of a conventional embedded microstrip transmission line (EMTL), with the length of 60 cm, while excited with rectangular pulses with different pulse widths (PW=50 ps, 450 ps, and 1 ns). The results are compared with sine wave excitation for three VWC values (10%, 20%, and 30%). In this simulation, the moist sandy dispersion model was extracted from data in [25] and then CST software is used to define a dispersive material in the simulation process as the moist soil.

According to Fig. 2(a-c), distortions are generated in the output wave when EMTL is excited with pulses. This degradation increases by reducing the pulse width. Moreover, at a higher VWC level, distortion increases, and the rising/falling edges of the pulses are not precisely detectable. However, in the single tone excitation technique using a sine wave, no distortion occurs (Fig. 2(d)). Therefore, since pulse distortion in dispersive and inhomogeneous media is a drawback of time-domain sensors, we propose a TDT-SMS based on the sine wave excitation technique to measure the complex permittivity of MUT (moist soil).

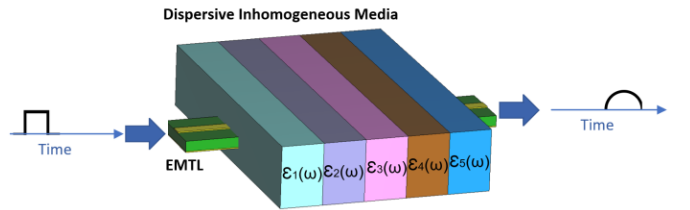

Fig. 1. Pulse distortion in a dispersive inhomogeneous media with five different materials ($\varepsilon_1(\omega), \varepsilon_2(\omega), \varepsilon_3(\omega), \varepsilon_4(\omega), \varepsilon_5(\omega)$).

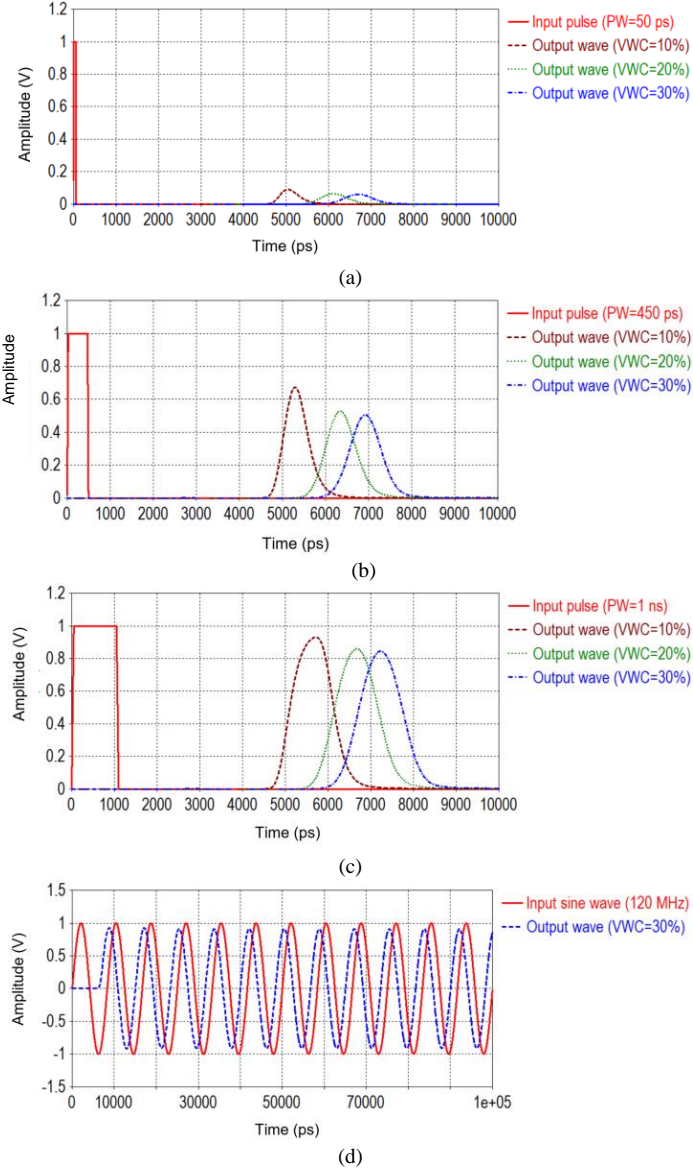

Fig. 2. Distortion in a dispersive media for pulse and sine wave excitations in different VWC values, a) PW=50 ps, b) PW=450 ps, c) PW=1 ns, d) 120 MHz sine wave.

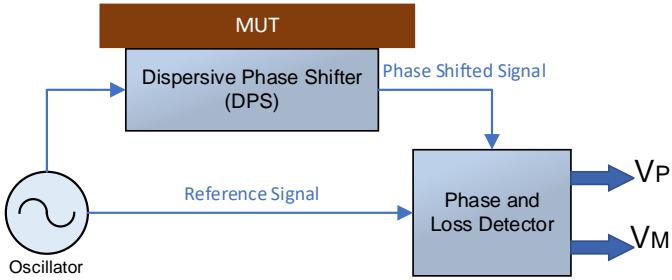

Fig. 3. Schematic of the proposed TDT-SMS based on sine wave excitation.

### B. Compact Dispersive Phase Shifter (DPS) Design

The proposed TDT-SMS consists of three main parts: reference oscillator, dispersive phase shifter (DPS), and phase/loss detector (Fig. 3). The phase and magnitude difference between the reference and phase-shifted signals are measured using a phase/loss detector, when the DPS is excited with a sine wave and surrounded by the moist soil (MUT). Measured phase and magnitude are related to $\varepsilon'$ and $\varepsilon''$, respectively, and the proposed sensor is capable to report the complex permittivity of moist soil (MUT). In this subsection, design theory of the proposed DPS is discussed.

There are two challenges of using a microstrip TL to realize a phase shifter in the proposed TDT-SMS under sine wave excitation. First, long TL is required to achieve an adequate phase shift, leading to a measurable phase difference between the output wave and reference signal. Second, the low sensitivity of conventional microstrip TL, $S_M$, which is the ratio of phase difference variation ($d(\Delta\emptyset)$) per unit variation of the MUT relative permittivity ($\varepsilon_{rm}$) as:

$$S_M = \frac{d(\Delta\emptyset)}{d(\varepsilon_{rm})} \quad (1)$$

For instance, conventional microstrip TL on FR-4 substrate with thickness of 0.6 mm exhibits 5.5 ps/mm time delay, where the phase difference at 120 MHZ is around $0.23^0$/mm. Therefore, a length of 10 cm is required to achieve $23^0$ phase differences (measurable value), which is not suitable for compact sensor design. Further, the $S_M$ for this microstrip TL is $0.023^0$/mm and a microstrip TL with length of 10 cm reaches $S_M = 23^0$. Therefore, we propose a TDT-SMS that uses a DPS to achieve high phase difference and high sensitivity in a compact structure.

The 3D layers schematic, equivalent circuit model, and layers of the proposed DPS are presented in Fig. 4. This structure consists of three layers: interdigital capacitor and conventional microstrip line (top layer), 4-CSR (middle layer), and ground metal cover (bottom layer), and between these metal layers are filled with the same substrate (FR-4). The middle and bottom layers form a stacked 4-CSR (S4-CSR) which improves the resolution of the proposed TDT-SMS and its theory is discussed in this section.

Due to the small electrical dimensions of DPS at the resonance frequency, it can be described using lumped element equivalent circuits. In the equivalent circuit in Fig. 4(b), $L$ and $C_u$ indicate the per-section inductance and capacitance of the top layer, $C_c$ and $L_c$ model the 4-CSR, $C_d$ represents the capacitance between middle and bottom layers, and $C_i$ denotes the series interdigital capacitance on the top layer.

The dispersion relation of the DPS is deduced from the equivalent circuit as [26]:

$$\cos(\beta l) = 1 + ZY/2 \quad (2)$$

$$Z_c = \sqrt{\frac{Z}{2}\left(\frac{Z}{2} + \frac{2}{Y}\right)} \quad (3)$$

The allowed band for backward-wave propagation in the structures occurs in the region where the characteristic impedance, $Z_c$, and the phase constant, $\beta l$ (given by expressions (1) and (2)) are both real numbers. From [27] and considering $C_t = C_u + C_d$ as the total capacitance between layers, the upper and lower cutoff frequency bands, $f_{cu}$ and $f_{cl}$, are given as:

$$f_{cu} = \frac{1}{2\pi\sqrt{LC_i}}, \quad f_{cl} = \frac{\sqrt{b - \sqrt{b^2 - 4ac}}}{2\pi\sqrt{2a}} \quad (4)$$

Where

$$\begin{cases} a = CLL_cC_cC_i \\ b = CLC_i + 8C_iL_c(C_c + C) + L_cC_c \\ c = 8C_i + C \end{cases} \quad (5)$$

Moreover, according to the equivalent circuit model in Fig. 4(b), there are two transmission zero frequencies at $Z = \infty$ and $Y = \infty$ as:

$$Z = \infty \rightarrow \frac{j\omega L}{2} + \frac{1}{2j\omega C_i} = \infty \rightarrow f_{z1} = 0 \quad (6)$$

and

$$Y = \infty \rightarrow \frac{j\omega C_t(1-\omega^2 L_c C_c)}{1-\omega^2 L_c(C_c+C_t)} = \infty \rightarrow f_{z2} = \frac{1}{2\pi\sqrt{L_c(C_c+C_t)}} \quad (7)$$

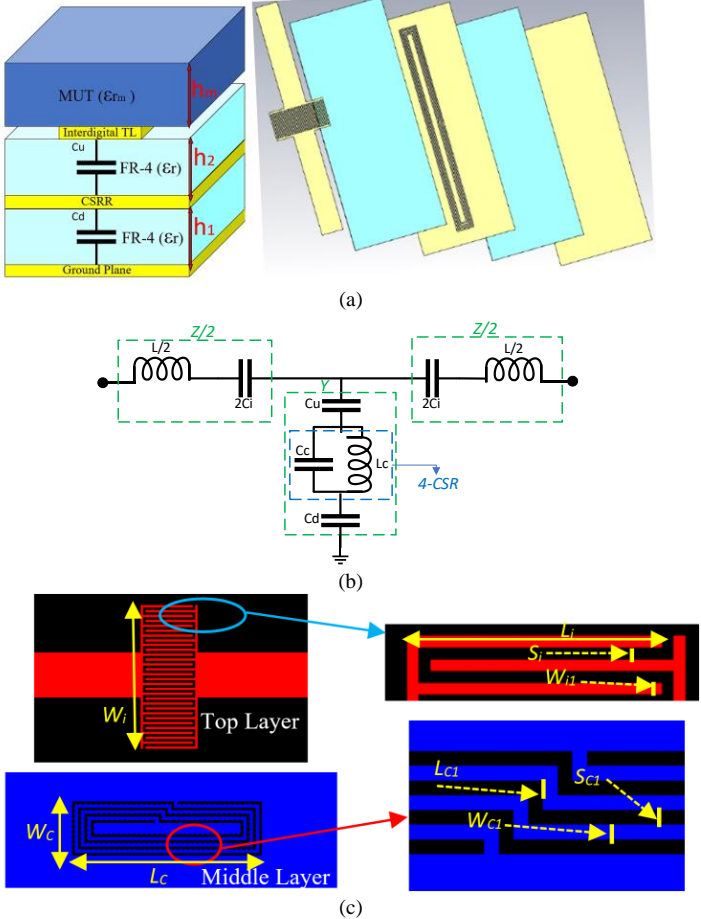

Fig. 4. Proposed TDT-SMS (a) 3D view of layers, (b) equivalent circuit model, (c) top and middle layers.

$f_{z1} = 0$ is far from the operational frequency band. However, $f_{z2}$ can be considered very close to the lower cutoff frequency band ($f_{cl}$) to achieve more rejection in the out of band region. In addition, in section III, it is shown that choosing $f_{z2}$ very close to $f_{cl}$ improves the sensitivity of TDT-SMS relative to changing of soil permittivity.

*C. TDT-SMS Sensitivity Calculation Approach*

Now, the effect of each component in the equivalent circuit model of the proposed DPS (Fig. 4(b)) on the sensing procedure is investigated. The inductors in the DPS equivalent circuit model ($L$, $L_c$) are not impacted by the MUT permittivity, and they are approximately constant in the measurement procedure. Therefore, capacitances ($C_u$, $C_d$, $C_c$, $C_i$) are key components in the DPS analysis.

The capacitance $C_c$ in Fig. 4(b), corresponds to the 4-CSR structure in the middle layer. Total equivalent capacitance of the slot structure, which is the sum of series combination of capacitances in each turn for 4-CSR structure can be calculated from [28]:

$$\frac{1}{C_c} = \sum_{n=1}^{N} \frac{1}{\varepsilon_0\left(\frac{\varepsilon_r+1}{2}\right)\left[\frac{P_n t_m}{S_c} + \frac{2\pi t_m}{\ln(8t_m/P_n)}\right]} \quad (8)$$

Where $t_m$, $P_n$ and $S_c$ are metal thickness, CSRs lengths, and gap width, respectively. According to (8), $C_c$ is not related to the MUT permittivity, and it can be ignored in the sensitivity calculation of the proposed TDT-SMS.

Moreover, the calculation of $C_u$ and $C_d$ is more straightforward and can be defined as the capacitance between two metal parallel plates (Fig. 4(a)). In $C_d$, if the narrow slots on the middle layer (4-CSR layer) are ignored, the bottom substrate is approximately surrounded between two metal plates (4-CSR layer and ground plane) and isolated from MUT and $\varepsilon_m$ variations. Moreover, the simulated and measured results in section III confirm the assumption of ignoring narrow slots on the middle layer in $C_d$ calculation.

However, in $C_u$ (Fig. 4(a)), the upper substrate is limited between the middle layer and interdigital/microstrip TL structure on the top layer which is embedded inside a moist soil (MUT). In this case, there are fringing fields on the top layer, and the effective area of the top plate changes relative to $\varepsilon_m$ variation. Therefore, the values of $C_d$ and $C_u$ can be calculated as [29]:

$$C_d = \varepsilon_0 \varepsilon_r \frac{A_d}{h_d} \quad (9)$$

$$C_u = \varepsilon_0 \varepsilon_r \frac{A'_u}{h_u} \quad (10)$$

Where $A_d$, $h_d$, and $h_u$ are physical area of the bottom ground layer, thicknesses of the lower and upper substrates, respectively, and:

$$A'_u = a \cdot (b + 2\Delta L) = A_u + 2a\Delta L \quad (11)$$

Where $a$, $b$, $A'_u$ are physical width, length, and effective area of the top layer, respectively, considering effective length increment, $\Delta L$, as [29]:

$$\Delta L = 0.412h \left[\frac{(\varepsilon_{reff}+0.3)(h_u+0.3)}{(\varepsilon_{reff}-0.258)(a/h_u+0.8)}\right] \quad (12)$$

Effective permittivity ($\varepsilon_{reff}$) of a microstrip line that is buried under a MUT can be expressed as [30]:

$$\varepsilon_{eff} = \varepsilon_0 \left[(\varepsilon_r + \varepsilon_m)/2 + \left((\varepsilon_r - \varepsilon_m)/2\sqrt{1 + 12\frac{h_u}{a}}\right)\right] \text{ if } h_m \gg h_u \quad (13)$$

Where $h_m$ is MUT thickness. Therefore, according to (9)-(13), $C_d$ and $C_u$ are calculated and the impact of MUT permittivity ($\varepsilon_m$) on them is considered.

The fourth capacitance in the SMS equivalent circuit model is $C_i$ which is related to the interdigital capacitance on the top layer. Several formulas to calculate this capacitance are presented and we used the following one [31]:

$$C_i(pF) = (\varepsilon_{reff} + 1)l_i[(N-3)A_1 + A_2] \quad (14)$$

Where

$$A_1 = 4.409 \tanh\left[0.55\left(\frac{h_u}{W_i}\right)^{0.45}\right], A_2 = 9.92 \tanh\left[0.52\left(\frac{h_u}{W_i}\right)^{0.5}\right] \quad (15)$$

Finally, according to the investigation of capacitances in the equivalent circuit model (Fig. 4(b)), sensitivity of the proposed DPS ($S_{DPS}$) relative to $\varepsilon_{rm}$ variation is derived as:

$$S_{DPS} = \frac{d(\Delta\emptyset)}{d\varepsilon_{rm}} = \left(\frac{d(\Delta\emptyset)}{dC_u} \times \frac{dC_u}{d\varepsilon_{rm}}\right) + \left(\frac{d(\Delta\emptyset)}{dC_i} \times \frac{dC_i}{d\varepsilon_{rm}}\right) \quad (16)$$

As mentioned in this section, $\varepsilon_{rm}$ variation cannot impact the other components in the equivalent circuit model in Fig. 4(b) ($C_d$, $C_c$, $L_c$, $L$). Hence, we ignored these components in the sensitivity equation (15). In addition, the analytical calculation of equation (15) is complicated, so we use numerical solutions in Matlab.

## III. SIMULATION, MEASUREMENT, AND DISCUSSION

To demonstrate the functionality of the proposed TDT_SMS, DPS is designed, simulated, and optimized using CST simulator. Then, DPS is fabricated on low-cost FR-4 substrate with $\varepsilon_r = 4.3$, thickness of $h_1 = h_2 = 0.6$ mm, and a dissipation factor of 0.0037 at 120 MHz. Copper thicknesses of 35 µm were created for printing top, middle, and bottom metal layers (Fig. 5). According to the design procedure discussed in Section II, the proposed TDT-SMS dimension is derived (Table I), optimum values of the DPS equivalent circuit model in the unloaded state are calculated (Table II), and the layout is extracted using equations in [30].

The S (scattering)-parameters allow to accurately describe the input-output relationships between ports, and hence the properties of the proposed DPS as a 2-port network [31]. Theoretical, simulated, and measured S21 results of the unloaded DPS as a two-port network are presented in Fig. 6. which are in a great agreement. DPS exhibits in-band insertion loss of around 3 dB, and 3-dB passband range from 114 MHz to 135 MHz. As mentioned in Section II, there is a zero-transfer frequency near the lower cutoff frequency band at 103 MHz.

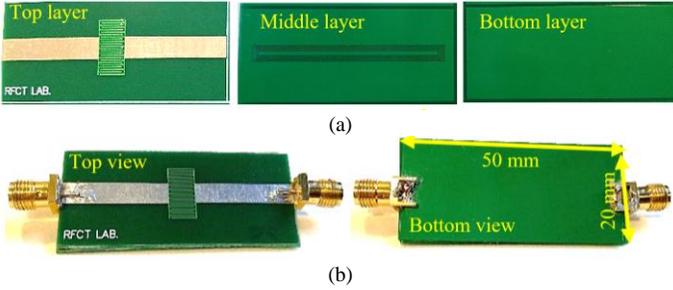

Fig. 5. Prototype of the proposed DPS (a) top layer, middle layer and bottom layer, (b) assembled DPS (top and bottom view).

TABLE I. DIMENSION OF THE DPS.

| $L_i$ (mm) | $W_i$ (mm) | $S_i$ (mm) | $W_{i1}$ (mm) | $L_C$ (mm) | $W_C$ (mm) | $L_{C1}$ (mm) | $S_{C1}$ (mm) | $W_{C1}$ (mm) |
|---|---|---|---|---|---|---|---|---|
| 5.6 | 17.4 | 0.2 | 0.2 | 44.5 | 3.7 | 0.2 | 0.2 | 0.2 |

TABLE II. VALUES OF THE EQUIVALENT CIRCUIT MODEL OF DPS.

| $L$ (nH) | $C_i$ (pF) | $C_u$ (pF) | $C_d$ (pF) | $C_c$ (pF) | $L_c$ (nH) |
|---|---|---|---|---|---|
| 1.5 | 1.2 | 14.1 | 15.9 | 7.8 | 17.2 |

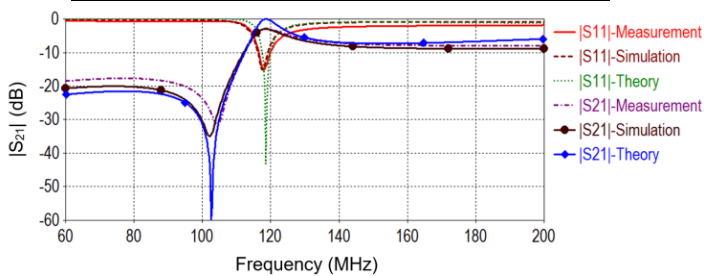

(a)

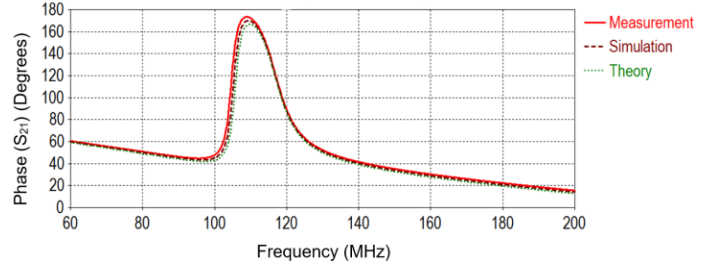

(b)

Fig. 6. a) Theoretical, simulated, and measured results of unloaded sensor ($\varepsilon_m = 1$), a) $|S_{11}|$ and $|S_{21}|$, b) phase($S_{21}$), vs frequency.

### A. DPS Simulation and Measurement Results vs VWC

Several relations between soil moisture content and soil dielectric constant have been proposed. Table III presents the real and imaginary parts of the dielectric constant values as a function of VWC around 130 MHz [25]. After validating the proposed sensor performance in the unloaded state (Fig. 6), the structure has been embedded into the sandy soil with different VWC. Different soil VWC levels have been achieved based on that has been released by the Department of Sustainable Natural Resources NSW Australia [25]. Firstly, the sand has been put in the oven to dry it. Then, the water has been added to provide sand with 5%, 10%, 15%, 20%, 25%, and 30% VWC levels as:

$$VWC(\%) = \frac{W_2}{W_1+W_2} \times 100 \quad (17)$$

Where W1 and W2 are the weights of dried soil and added water, respectively. Furthermore, we measured performance of the proposed sensor in 5 trials for each VWC value.

The DPS was measured using a Vector Network Analyzer (VNA-ZVA40). Figure 7 shows the measured S-parameters results of the proposed DPS versus sandy soil with VWC of 0%, 10%, 20%, and 30%. According to Fig. 7(a), the proposed sensor operates as a band-pass filter over 114 MHz to 125 MHz. By increasing the VWC value of MUT from 0% to 30%, zero frequency, lower and upper cutoff frequency bands increase. Further, the measured insertion loss of the DPS at the operational frequency bands is better than 3 dB. Figure 7(b) demonstrates the measured results of phase difference (phase shift). It is demonstrated that each phase difference curve, which is related to a specific VWC value, has a unique pattern. This property is used to measure the VWC of soil in Subsection *C*.

TABLE III. DIELECTRIC CONSTANT VALUES AS A FUNCTION OF VOLUMETRIC WATER CONTENT FOR SAND AROUND 130 MHz [24]

| VWC (%) | 0 | 5 | 10 | 15 | 20 | 25 | 30 |
|---|---|---|---|---|---|---|---|
| $\varepsilon'_m$ | 2.5 | 6 | 8 | 14.5 | 18 | 21 | 23.5 |
| $\varepsilon''_m$ | 0.05 | 0.5 | 0.9 | 1.8 | 2.5 | 3.1 | 3.5 |

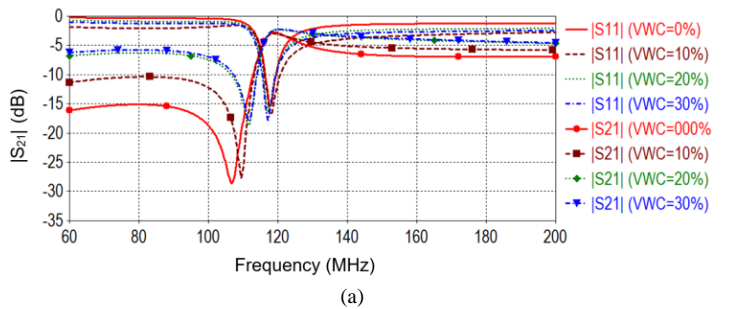

(a)

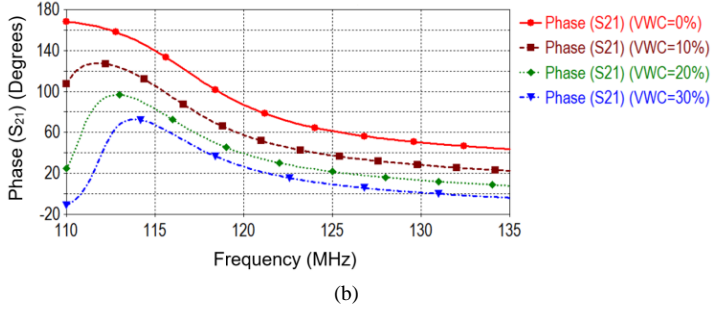

(b)

Fig. 7. Measurement results of the proposed DPS at four different values of VWC (0%, 10%, 20%, 30%) for sandy soil, a) |S$_{11}$| and |S$_{21}$|, b) phase difference (phase shift).

*B. DPS Sensitivity Analysis Results*

In this subsection, the sensitivity simulation results of the proposed DPS that was investigated in Section *II.B*, are presented. Effective length increment, $\Delta L$, of the top layer vs the MUT permittivity is presented in Fig. 8(a), while DPS is embedded in MUT. As it is shown, $\Delta L$ decreases from 270 µm to 250 µm due to $\varepsilon_{rm}$ variation (1 to 24). Figure 8(b) exhibits theoretical calculation results of $C_u$ and $C_i$ versus the MUT permittivity ($\varepsilon_{rm}$), based on effective $\Delta L$ value (equations (9) and (13)). According to this figure, when $\varepsilon_{rm}$ changes from 1 to 24, maximum variations of $C_u$ and $C_i$ are 15 pF and 1 pF, respectively. Hence, $C_u$ is the most important component in the DPS system to achieve highly sensitive TDT-SMS.

Figure 9 (a) presents simulated |S$_{21}$| of the proposed DPS vs frequency and MUT permittivity. DPS insertion loss is approximately 3 dB in the passband range when $\varepsilon_{rm}$ changes from 1 to 24. Furthermore, rejection at the zero-transfer frequency (near the lower cutoff frequency band) decreases as MUT permittivity increases.

Now, in the sine wave (single tone) excitation scenario, we find an optimum frequency in the passband and excite the DPS at this frequency to achieve high sensitivity, related to the MUT permittivity variation. In this regard, according to sensitivity equation (15), Fig. 9(b) presents the numerical solution of this equation in Matlab software. This figure shows that, at a fixed operational frequency band, the DPS sensitivity ($S_{DPS}$) decreases as MUT permittivity increases, while the maximum sensitivity occurs around the lower cutoff frequency band of DPS. For instance, the sensitivity values at 114 MHz for $\varepsilon_{rm} = 5, 10, 15,$ and 20 are around $30^0$, $16^0$, $8^0$, and $7^0$, respectively. Therefore, we select 114 MHz as the optimum frequency of the sine wave in the single tone excitation scenario.

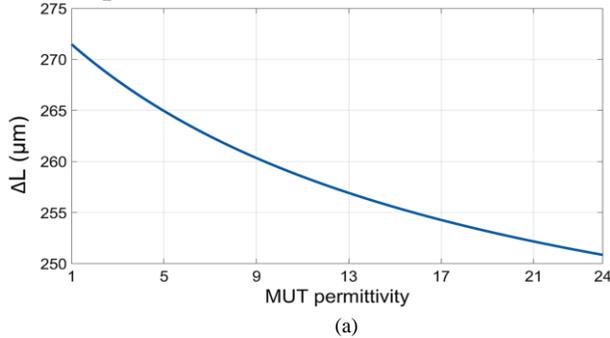

(a)

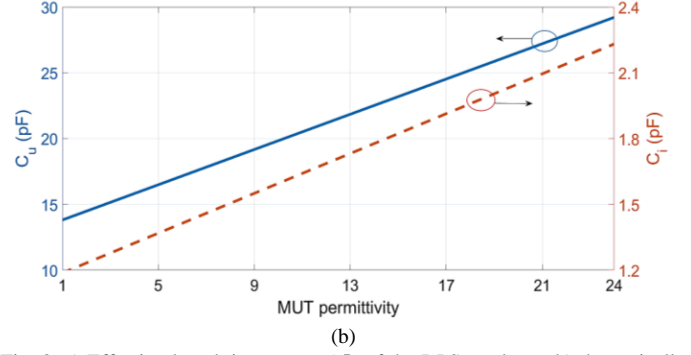

(b)

Fig. 8. a) Effective length increment, $\Delta L$, of the DPS top layer, b) theoretically calculated values of $C_u$ and $C_i$ capacitances, vs MUT permittivity ($\varepsilon_{rm}$).

According to Fig. 9(b), in the worst-case scenario the DPS sensitivity at $\varepsilon_m = 23.5$ (VWC = 30%) is $S_{DPS} = \frac{d(\Delta\emptyset)}{d\varepsilon_{rm}} = 7^0$ and based on the measured data in [25], $\frac{d(VWC\%)}{d\varepsilon_{rm}} = 2$. Hence, for the proposed DPS, the phase difference variation related to VWC is calculated as:

$$\frac{d(\Delta\emptyset)}{d(VWC\%)} = 3.5^0 \quad (18)$$

Therefore, according to (18), $\Delta\emptyset = 3.5^0$ must be detectable using phase/loss detector to achieve 1% resolution in VWC. This total resolution of the proposed technique is investigated in Section III (c).

*C. Embedded Realization of TDT-SMS*

In the proposed TDT-SMS, phase and amplitude differences between the reference and phase-shifted signals are detected using a phase/loss detector (Fig. 3). The measured phase and amplitude differences are related to the real and imaginary parts of MUT permittivity, respectively. In the embedded scenario, the detector converts these differences to output voltages $V_P$ and $V_M$ which are related to phase differences and loss, respectively, and can be measured using a simple multimeter. We used AD8302 which is a fully integrated system for measuring phase and loss [32]. Two outputs of AD8302, $V_M$ and $V_P$, provide an accurate measurement of loss over a ±30 dB range scaled to 30 mV/dB, and phase over a 0°–180° range scaled to 10 mV/degree. According to the AD8302 datasheet, at 100 MHz, the accuracy of measuring phase difference, from $30^0$ to $150^0$, is better than $0.1^0$ and the accuracy of loss measurement from -25 dB to 0 dB, is 0.1 dB [32]. Therefore, the loss and phase measurement region should be considered in this area to achieve a high accuracy sensor.

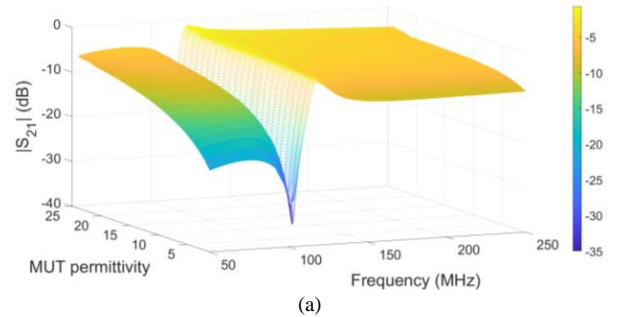

(a)

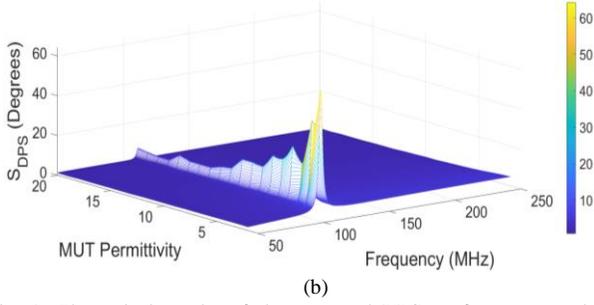

(b)

Fig. 9. Theoretical results of the proposed DPS vs frequency and MUT permittivity ($\varepsilon_{rm}$), a) $|S_{21}|$, b) DPS sensitivity ($S_{DPS}$).

Figure 10(a) shows the block diagram of the measurement setup. A 2-channel Vector Signal Generator (SMW200A) is used as the reference signal generator. In order to verify the sensor accuracy, an Oscilloscope (RTO2044) is used to measure phase-shifted and reference signals and verify the $V_P$ and $V_M$ results, which were measured using AD8302 and Multimeter (HMC8012). Finally, we used two power splitters (ZFSC-2-4) to organize measuring and verification procedures, simultaneously. Figures 10(b), (c), (d) show the measurement setup in unloaded, dry soil, and moist soil cases, respectively. Before starting the measurement process, the setup is calibrated considering the loss and phase differences of all signal directions.

In the proposed sensing method, the VWC measurement is based on the permittivity detection technique of moist soil and then relating the measured permittivity to VWC of soil. Therefore, the resolution of VWC measurement depends on the permittivity measurement resolution. According to Fig. 10(a), since the phase difference of DPS output signal is measured relative to the reference oscillator using phase/loss detector, the AD8302 resolution determines the resolution of testing parameter. From (18), the accuracy of measuring phase difference using AD8302 is $0.1^0$, and the total resolution of the proposed sensing technique is:

$$\Delta\emptyset > 0.1^0 \rightarrow \Delta(VWC\%) = 0.03\% \quad (19)$$

However, 0.03% is a theoretical value for resolution and in the real measurement setup, there are several noise sources and interference that can degrade the resolution value. According to the commercialized soil moisture sensors [33], VWC=1% is an acceptable value and the proposed sensor meets this amount.

There are three stages of testing permittivity and VWC in the proposed technique as shown in Fig. 11. According to Table III and existing data in [25], nominal values are provided for real and imaginary parts of soil permittivity relevant to each VWC. At this stage, we assigned the nominal values of permittivity ($\varepsilon_{mn}$) and volumetric water content ($(VWC)_n$) to the produced moist soil. Then, the measured values of soil permittivity are compared with the nominal values and estimation error is defined as:

$$Estimation\ error = \frac{|\varepsilon_m - \varepsilon_{mn}|}{\varepsilon_{mn}} \times 100 \quad (20)$$

Figure 12 demonstrates the measured results of testing setup in Fig. 10 when TDT-SMS was excited with 114 MHz sine wave (the results were recorded in five trials). Figure 12(a) and (b) show the extracted real and imaginary parts of the MUT (sandy soil) in VWC of 0%, 5%, 10%, 15%, 20%, 25%, and 30%. These figures exhibit an average error of the measured real and imaginary parts 3.5% and 11%, respectively. After measuring real and imaginary parts of soil permittivity, Fig. 12(a) and Fig. 12(b), the soil moisture value (VWC) is calculated using data in Table III. According to Fig. 12(c), the estimation error is less than ±0.3%, ±0.6%, ±0.8%, and ±1.2% at 0%, 10%, 20%, and 30% VWC of sandy soil, respectively. The achieved measurement results confirm the accuracy of the proposed sensing technique and design procedure.

Table IV compares the performance of the proposed sensor with other reported sensors. Moreover, to provide a fair comparison with previously published papers in this table, the references are compared with the proposed technique in terms of permittivity which is a general parameter to compare different sensors, regardless of their applications. Furthermore, in terms of different compositions of soil and its effect on the permittivity measurement accuracy, we consider sandy soil type in our study so that the permittivity of various dried sandy soil occupied a narrow range of values (less than 0.3 variation in the permittivity) around operating frequency range (114 MHz) and it confirms the reasonability of the comparison process.

From Table IV, it is evident that, with 3.5% measured error of the real permittivity, the proposed sensor exhibits higher accuracy in a broad range of permittivity (1 to 23.5) in comparison with other works. Moreover, most of reported works are based on FDR technique which has several challenges in real applications and field measurements. This comparison proves the usefulness of our proposed TDT-SMS sensor for precision farming applications.

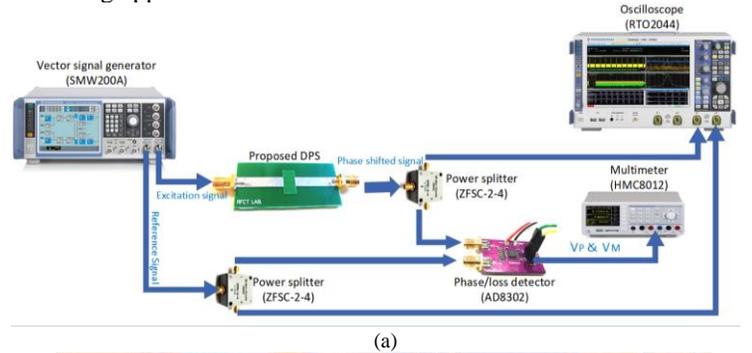

(a)

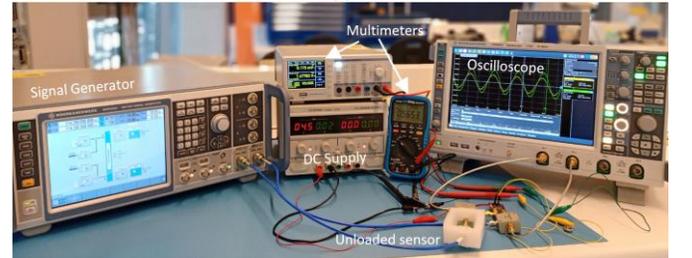

(b)

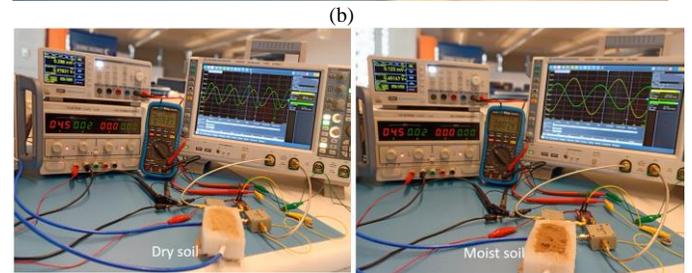

(c) (d)

Fig. 10. a) Measurement setup of embedded TDT-SMS and required equipment, b) unloaded, c) dry soil, d) moist soil.

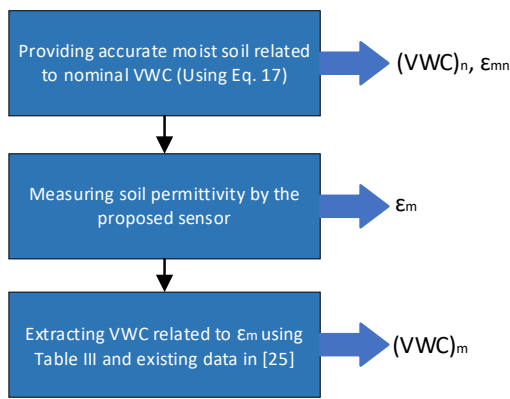

Fig. 11. Testing process of permittivity and VWC for moist soil.

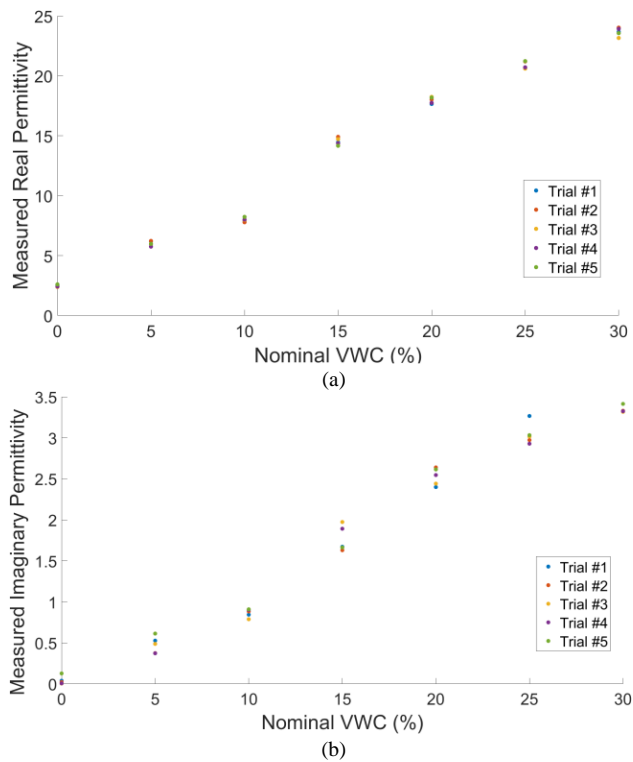

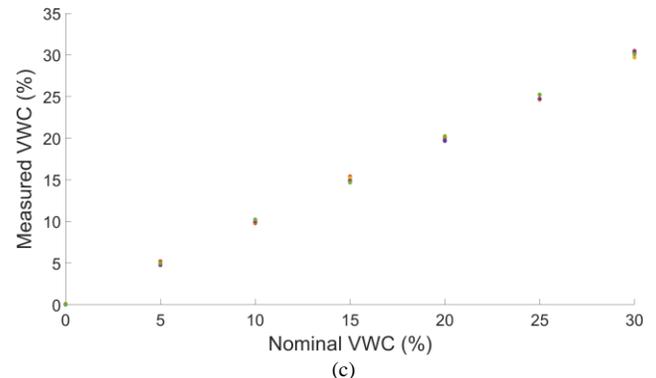

Fig. 12. Measured results of sandy soil for different nominal VWC values in five trials, a) real permittivity, b) imaginary permittivity, c) VWC.

## IV. CONCLUSION

This work demonstrated the design, simulation, and measurement of a highly sensitive and compact time-domain transmissometry soil moisture sensor (TDT-SMS) based on a dispersive phase shifter (DPS). The proposed sensor achieved high phase shift value, low profile, low cost, simple usage and high sensitivity for different VWC values. To achieve a low profile DPS, an interdigital capacitor was used on the top layer and stacked 4-CSR on the bottom layer of the proposed DPS. The TDT-SMS is excited with a 114 MHz sine wave as the reference oscillator. By measuring the phase difference and loss between the reference and phase-shifted signals, real and imaginary parts of the soil permittivity were measured, respectively, and VWC was extracted. Accuracy of ±1.2% at VWC of 30% was achieved (worst-case scenario for sensitivity), which is very suitable in sensing applications. Furthermore, the developed design guideline of TDT-SMS can be used in the measuring process of different materials under test (MUTs). Hence, the proposed time-domain technique can be adopted to different scenarios and applications. It is the scope of our future work to design an embedded low power soil moisture sensor and eliminate any need for testing equipment. In addition, Energy Harvester (EH) or Wireless Power Transfer (WPT) devices [22], [33], [34], can be integrated with the proposed sensor to realize a self-sustainable sensor ($S^3$) that is of paramount importance in field measurements for precision agriculture and industrial applications.

TABLE IV. COMPARISON TABLE

| Ref. | Technique | Frequency Band (GHz) | Sensitivity in Permittivity Measurement | Measured Permittivity Range | Permittivity Measurement Error (%) | Application |
|---|---|---|---|---|---|---|
| [35] | FDR | 1.7 | 33.3 MHz | 1~10.2 | NA | Comparator functionality |
| [36] | FDR | 4.5 | 60 MHz | 4.4 | 4.3 | NA |
| [37] | FDR | 2.7 | NA | 10.2 | 7.6 | NA |
| [38] | FDR | 5.5 | 3.34 MHz | 30 | NA | Microfluid |
| [11] | TDR (Pulse) | 0.01~1 | NA | 1~20 (real), 1~7 (imaginary) | 5 (non dispersive), 20 (dispersive) | Soil moisture |
| This Work | TDT (Single tone) | 0.114 | $7^0$ | 1~23.5 (real), 0.1~4 (imaginary) | 3.5 (non dispersive), 3.5 (dispersive) | Soil moisture |

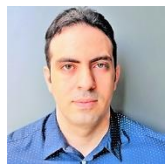

**Rasool Keshavarz** was born in Shiraz, Iran in 1986. He received the Ph.D. degree in Telecommunications Engineering from the Amirkabir University of Technology, Tehran, Iran in 2017 and is currently working as Postdoctoral Research Associate in RFCT Lab at the University of Technology, Sydney, Australia. His main research interests are RF and microwave circuit and system design, sensors, antenna design, wireless power transfer (WPT), and RF energy harvesting (EH).

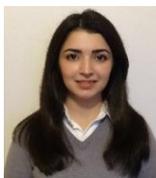

**Negin Shariati** is a Senior Lecturer in the School of Electrical and Data Engineering, Faculty of Engineering and IT, University of Technology Sydney (UTS), Australia. She established the state of the art RF and Communication Technologies (RFCT) research laboratory at UTS in 2018, where she is currently the Co-Director and leads research and development in RF-Electronics, Sustainable Sensing, Low-power Internet of Things, and Energy Harvesting. She leads the Sensing Innovations Constellation at Food Agility CRC (Corporative Research Centre), enabling new innovations in agriculture technologies by focusing on three key interrelated streams; Sensing, Power and Connectivity. Since 2018, she has held a joint appointment as a Senior Lecturer at Hokkaido University, externally engaging with research and teaching activities in Japan. She attracted over one million dollars worth of research funding across a number of CRC and industry projects, where she has taken the lead CI role and also contributed as a member of the CI team. Negin Shariati completed her PhD in Electrical-Electronic and Communication Technologies at Royal Melbourne Institute of Technology (RMIT), Australia, in 2016. She worked in industry as an Electrical-Electronic Engineer from 2009-2012. Her research interests are in Microwave Circuits and Systems, RF Energy Harvesting, low-power IoT, Simultaneous Wireless Information and Power Transfer, Wireless Sensor Networks, Antenna and AgTech.

## V. ABBREVIATION

A glossary of used terms is given below:

| Acronym | Description |
|---|---|
| CSRR | Complement of split-ring resonator |
| DPS | Dispersive phase shifter |
| EH | Energy harvester |
| EMTL | Embedded microstrip transmission line |
| FDR | Frequency domain reflectometry |
| IoT | Internet of Things |
| MUT | Material under test |
| PW | Pulse width |
| $S_{11}$ | The reflection coefficient between the port impedance and the network's input impedance at Port 1 |
| $S_{21}$ | Represents the power transferred from Port 1 to Port 2 of a two-port network |
| S4-CSR | Stacked 4-turn Complementary Spiral Resonator |
| SRR | Split-ring resonator |
| TDR | Time-domain reflectometry |
| TDT-SMS | Time-domain transmissometry soil moisture sensor |
| VNA | Vector network analyzer |
| VUT | Volume under test |
| VWC | Volumetric water content |
| WPT | Wireless power transfer |
| WSN | Wireless sensor networks |